# Superconductivity in $Li_3Ca_2C_6$ intercalated graphite

Nicolas Emery[1], Claire Hérold*[1], Jean-François Marêché[1], Christine Bellouard[2], Geneviève Loupias[3] and Philippe Lagrange[1,4]

1) Laboratoire de Chimie du Solide Minéral (UMR CNRS 7555), Université Henri Poincaré Nancy I, B.P. 239, 54506 – Vandoeuvre-lès-Nancy Cedex, France
2) Laboratoire de Physique des Matériaux (UMR CNRS 7556), Université Henri Poincaré Nancy I, B.P. 239, 54506 – Vandoeuvre-lès-Nancy Cedex, France
3) Institut de Minéralogie et de Physique des Milieux Condensés (UMR CNRS 7590), Université Pierre et Marie Curie (Paris 6), case 115, 4 place Jussieu, 75252 Paris Cedex 05, France
4) Ecole Européenne d'Ingénieurs en Génie des Matériaux, Institut National Polytechnique de Lorraine, 6 rue Bastien Lepage, B.P. 630, 54010 – Nancy Cedex, France

Correspondence should be addressed to C. H. (Claire.Herold@lcsm.uhp-nancy.fr).

In this letter, we report the discovery of superconductivity in $Li_3Ca_2C_6$. Several graphite intercalation compounds (GICs) with electron donors, are well known as superconductors[1]. It is probably not astonishing, since it is generally admitted that low dimensionality promotes high superconducting transition temperatures. Superconductivity is lacking in pristine graphite, but after charging the graphene planes by intercalation, its electronic properties change considerably and superconducting behaviour can appear. $Li_3Ca_2C_6$ is a ternary GIC[2], for which the intercalated sheets are very thick and poly-layered (five lithium layers and two calcium ones). It contains a great amount of metal (five metallic atoms for six carbon ones). Its critical temperature of 11.15 K is very close to that of $CaC_6$ GIC[3,4] (11.5 K). Both $CaC_6$ and $Li_3Ca_2C_6$ GICs possess currently the highest transition temperatures among all the GICs.

Since forty years, we know that $KC_8$ GIC becomes superconductor[5] at very low temperature (0.14 K[6]). More recently, metastable binary compounds obtained by high-pressure methods show higher critical temperatures : 1.9 K for $LiC_2$, 3 K for $KC_3$, 5 K for $NaC_2$[7] or 5.5 K for $KC_4$[8]. On the other hand, ternary GICs prove to be superconductors as $KHgC_8$ (1.90 K)[9], $KTl_{1.5}C_4$ (2.70 K)[10] or $CsBi_{0.5}C_4$ (4.05 K)[11]. But, very recently, two binary GICs have been shown to become superconductors at higher temperatures : $YbC_6$ exhibits a critical temperature, $T_c$, of 6.5 K[3], and in $CaC_6$, $T_c$ reaches 11.5 K[3,4].

Our recent works concerning the intercalation into graphite of lithium-calcium alloys[2] showed the existence of two ternary lithium-calcium GICs belonging to the first stage, denoted α and β. This method can be used also in order to synthesise[12] bulk $CaC_6$. In this binary, only one calcium plane is intercalated between two successive graphene layers. On the contrary, both α and β ternary GICs possess intercalated alloy sheets consisting of several atomic layers. Indeed, five layers are present in an intercalated sheet of α compound, according to the Li-Ca-Li-Ca-Li stacking and seven layers in a β compound sheet, with the Li-Ca-Li-Li-Li-Ca-Li sequence. Their repeat distances are 776 and 970 pm for α and β compounds respectively, instead of 335 pm in pristine graphite. The chemical formulas reveal differences of intercalated metal amounts: $Li_{0.5}Ca_3C_6$ for α phase, and $Li_3Ca_2C_6$ for β one. These formulas were obtained from data collected by nuclear microprobe techniques[13].

As reported in the following, only the β phase exhibits a superconducting behaviour. Consequently, we will give some structural details[2] about the β compound only. Since the **c**- axes of all the crystallites forming the HOPG platelet are parallel to each other, the $Li_3Ca_2C_6$ sample **c**-axis direction is well-defined. The resulting *00l* X-ray reflections are isolated (Fig. 1a). All of them can be identified as 00*l* ternary β compound reflections, confirming the sample high purity.

By Fourier transform of the *00l* structure factors, the **c**-axis electronic density profile of the compound was shown in Fig. 1b and compared to the calculated one from an atomic stacking model. Of course, we searched to fit at best both profiles, taking chemical formula into account. Thus, the seven atomic layers present in the $Li_3Ca_2C_6$ intercalated sheet appear clearly. The three central lithium layers are flanked by two calcium ones and in turn by two more lithium layers.

While the sample **c**-axis is well-defined, the crystallites forming HOPG platelets are disordered in perpendicular directions, leading to an average of **a** and **b** directions, denoted as *ab*. The $Li_3Ca_2C_6$ *hk0* reflection study exhibits an hexagonal 2D unit cell, commensurate with that of graphite, and an « a » parameter of 745 pm, three times larger than the graphitic one. Nevertheless, the atomic positions inside the cell are difficult to specify, so that the $Li_3Ca_2C_6$ space group is currently unknown.

Superconductivity does not appear for α compound above 2 K, while it is revealed at 11.15 K for β one. Fig. 2 shows the results concerning β compound for both ZFC and FC measurements. We observed a clear diamagnetic transition at 11.15 K under an applied field of 50 Oe. However, the saturation of diamagnetism is not reached down to 2 K.

The critical fields and the superconducting volume fraction were estimated from magnetization measurements versus applied field M(H) at several temperatures. Thanks to the 2D character of GICs, two directions were studied : - applied field parallel to **c**-axis (//c) or – perpendicular to **c**-axis (//ab). For these data, we have taken into account the demagnetization correction due to sample shapes. In this case, the field is corrected with the following relation $H = H_0 - 4\pi NM$ (CGS units), where $H_0$ is the applied field, N is the shape correction factor and M is the magnetization of the sample. For the applied field parallel to **c**-axis, the N factor has been estimated to 0.7-0.8, and for the applied field perpendicular to the **c**-axis, the correction is negligible. As usual, the penetration length of the magnetic field in the sample was supposed quite weak so that we have disregarded the corresponding correction.

The superconducting volume fraction, calculated from the M(H) measurements with H applied parallel to the **c**-axis, was evaluated at 55 ± 5 %.

The magnetic phase diagram for the β phase is presented in Fig. 3. This material appears as a type II superconductor. Indeed, the lower critical field $H_{c1}$, defined as the linearity break in the first part of the M(H) curve at a given temperature, is approximately the same for both geometries (with extrapolated values: $H_{c1//ab}(0)$ = 32 Oe and $H_{c1//c}(0)$ = 30 Oe). However, the upper critical field, $H_{c2}$, defined as the beginning of the second linear domain, exhibits a sizable anisotropy : we estimate zero temperature extrapolated values $H_{c2//ab}(0)$ and $H_{c2//c}(0)$ of about 3100 Oe and 2000 Oe respectively, leading to a ratio ($H_{c2//ab}$ / $H_{c2//c}$) closed to 1.5. A so small anisotropic ratio implies that superconducting state of this material is anisotropic but three dimensional in spite of its layered structure, as already noticed in $YbC_6$[3] and $CaC_6$[4] compounds.

Although the measurements were repeated, using several « crystallographic pure samples », the complexity of $Li_3Ca_2C_6$ structural building probably generates defects which not allow to reach the complete diamagnetic saturation. However, $Li_3Ca_2C_6$ unambiguously appears as a new superconductor below 11.15 K. Currently, this compound possesses the highest critical temperature among the ternary GICs.

**Methods**

We have shown that the ternary GICs containing lithium are extremely unusual and very difficult to synthesise[14]. Indeed, various lithium-calcium ternary GICs were prepared using a well established protocol consisting of several successive steps[15].

For the β compound, it is necessary, in a first time, to synthesise a lithium-calcium alloy, whose composition is $Ca_{0.25}Li_{0.75}$. It is obtained with extremely pure lithium and calcium reagents, which are molten and well-homogeneously mixed, using a furnace in a glove box containing a very pure argon atmosphere. In a second time, a highly oriented pyrolytic graphite (HOPG) platelet is introduced in the liquid alloy, using a tungsten sample holder, in order to maintain an excellent contact between graphite and alloy. The intercalation reaction is carried out during 10 days at 250°C for one sample and at 300 °C for the second sample in a stainless steel reactor under pure argon atmosphere (this temperature range needs to be strictly observed, because, if the reaction temperature increases until 350°C, the reaction product becomes pure $CaC_6$ instead of the pure β ternary GIC). In a third time, when the reaction is ended, the bulk sample is carefully extracted of the liquid alloy. Then a two-sided cleavage of the intercalated platelet is performed in order to avoid any alloy deposit and surface defects. Finally, the sample is packed for its subsequent study.

The intercalated sample exhibits a pale yellow colour and the characteristic metallic brightness, with a remarkably strong hardness. By X-ray diffraction, it is very easy to isolate the *00l* reflections of the sample in order to test its purity.

Magnetization of the α and β compounds is measured as a function of temperature and magnetic field, using a Quantum Design MPMS5 SQUID magnetometer. Because of their reactivity, the samples have to be kept in a closed silica cell, under helium atmosphere. The samples are cooled down to 2 K without applied magnetic field. A field of 50 Oe is then applied and the magnetization of the samples is measured between 2 K and 15 K by steps of 0.2 K (« zero field cooling » experiment (ZFC)). After, the samples are cooled down to 2 K under the same field used in ZFC measurements. Then magnetization measurements (« field cooling » experiment (FC)) are performed like previously.

**Acknowledgements**

We acknowledge Matteo d'Astuto, Christophe Bellin and Sohrab Rabii for their respective help in this work.

**References**

1. T. Enoki, S. Masatsugu & E. Morinobu, *Graphite Intercalation Compounds and Applications* (Oxford Univ. Press, Oxford, 2003).
2. S. Pruvost, C. Hérold, A. Hérold & P. Lagrange, Structural study of novel graphite-lithium-calcium intercalation compounds. *Eur. J. Inorg. Chem.* **8** 1661-1667 (2004).
3. T. E. Weller, M. Ellerby, S. S. Saxena, R. P. Smith & N. T. Skipper, Superconductivity in the intercalated graphite compounds $C_6Yb$ and $C_6Ca$. *Nat. Phys.* **1**, 39-41 (2005).
4. N. Emery, C. Hérold, M. D'Astuto, V. Garcia, Ch. Bellin, J. F. Marêché, P. Lagrange & G. Loupias, Superconductivity of bulk $CaC_6$. *Phys. Rev. Lett.* **95**, 087003 (2005).
5. N. B. Hannay, T. H. B. T. Matthias, K. Andres, P. Schmidt & D. MacNair, Superconductivity in graphitic compounds. *Phys. Rev. Lett*. **14**, 225-226 (1965).
6. Y. Koike, S.-I. Tanuma, H. Suematsu & K. Higuchi, Superconductivity in the graphite-potassium intercalation compound $C_8K$. *J. Phys. Chem. Solids* **41**, 1111-1118 (1980).
7. I. T. Belash, O. V. Zharikov & A. V. Pal'nichenko, Superconductivity of GIC with Li, Na and K. *Synth. Met.* **34**, 455-460 (1989).
8. V. A. Nalimova, V. V. Avdeev & K. N. Semenenko, New Alkali Metal- Graphite Intercalation Compounds at high pressures. *Mat. Sci. Forum* **91-93** 11.16 (1992).

9. L. A. Pendrys *et al.* Superconductivity of the graphite intercalation compounds KHgC$_8$ and RbHgC$_8$. *Sol. St. Comm.* **38**, 677-678 (1981).

10. R. A. Wachnik, L. A. Pendris, F. L. Vogel & P. Lagrange, Superconductivity of graphite intercalated with thallium alloys. *Sol. St. Comm.* **43**, 5-8 (1982).

11. E. McRae, J. F. Marêché, A. Bendriss-Rerhrhaye, P. Lagrange & M. Lelaurain, New ternary superconducting graphite intercalation compounds containing bismuth and a heavy alkali metal. *Ann. Phys.* **11**, 13-22 (1986).

12. N. Emery, S. Pruvost, C. Hérold & P. Lagrange, New kinetical and thermodynamical data concerning the intercalation of lithium and calcium into graphite. *J. Phys. Chem. Solids* (2006) in press.

13. S. Pruvost, P. Berger, C. Hérold & P. Lagrange, Nuclear microanalysis : An efficient tool to study intercalation compounds containing lithium. *Carbon* **42**, 2049-2056 (2004).

14. S. Pruvost, C. Hérold, A. Hérold & P. Lagrange, On the great difficulty of intercalating lithium with a second element into graphite. *Carbon* **41**, 1281-1289 (2003).

15. S. Pruvost, C. Hérold, A. Hérold, J.-F. Marêché & P. Lagrange, Un nouveau composé d'intercalation du graphite : une phase lamellaire graphite-lithium-calcium. *C. R. Chimie* **5**, 559-564 (2002).

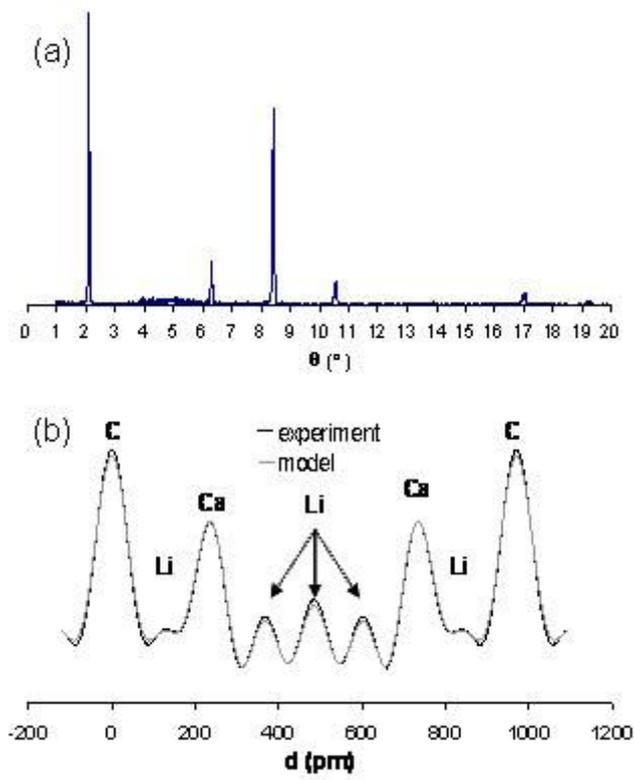

Fig. 1 : a) *00l* X-ray diffraction pattern for $Li_3Ca_2C_6$ (Mo $K_{\alpha 1}$), b) experimental and calculated **c**-axis electronic density profiles for $Li_3Ca_2C_6$.

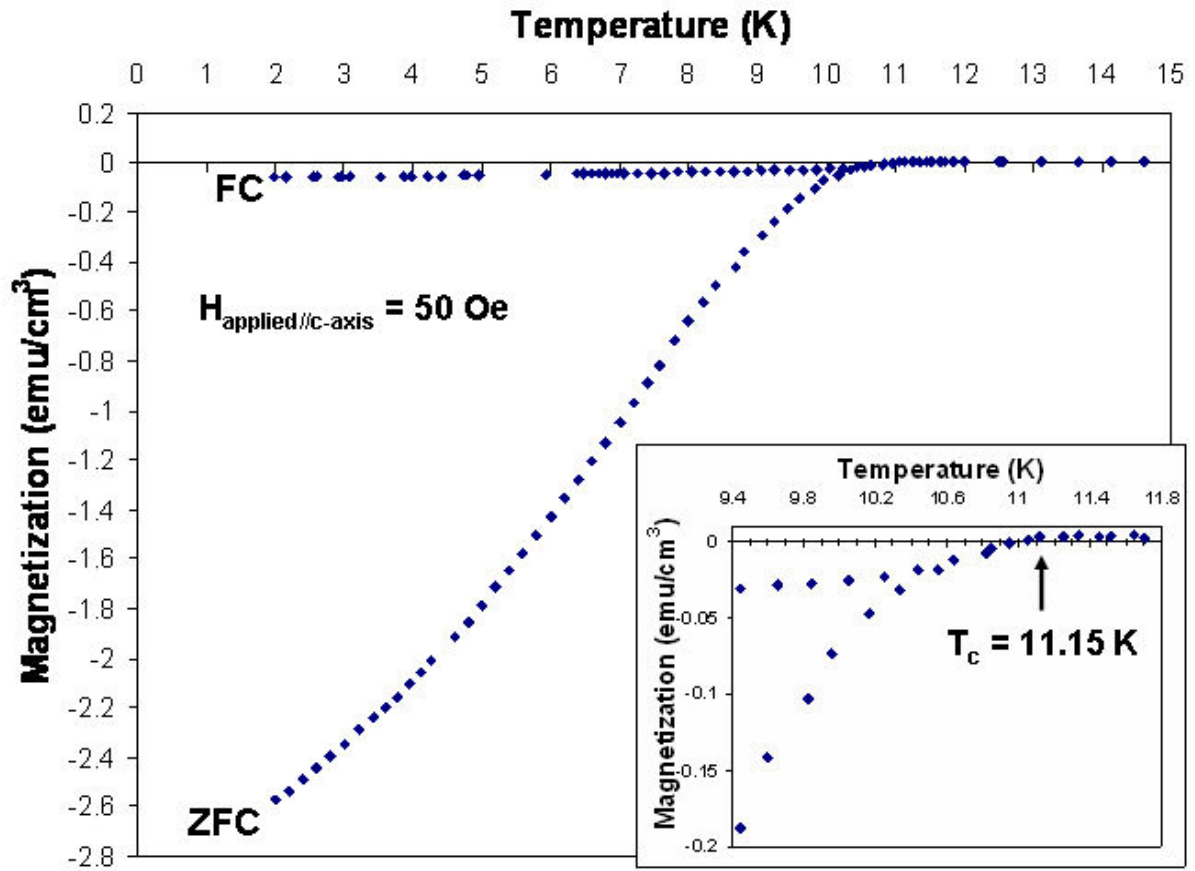

Fig. 2 : Magnetization of $Li_3Ca_2C_6$ versus temperature with a 50 Oe field applied parallel to the **c**-axis (field cooling FC, and zero field cooling ZFC measurements).

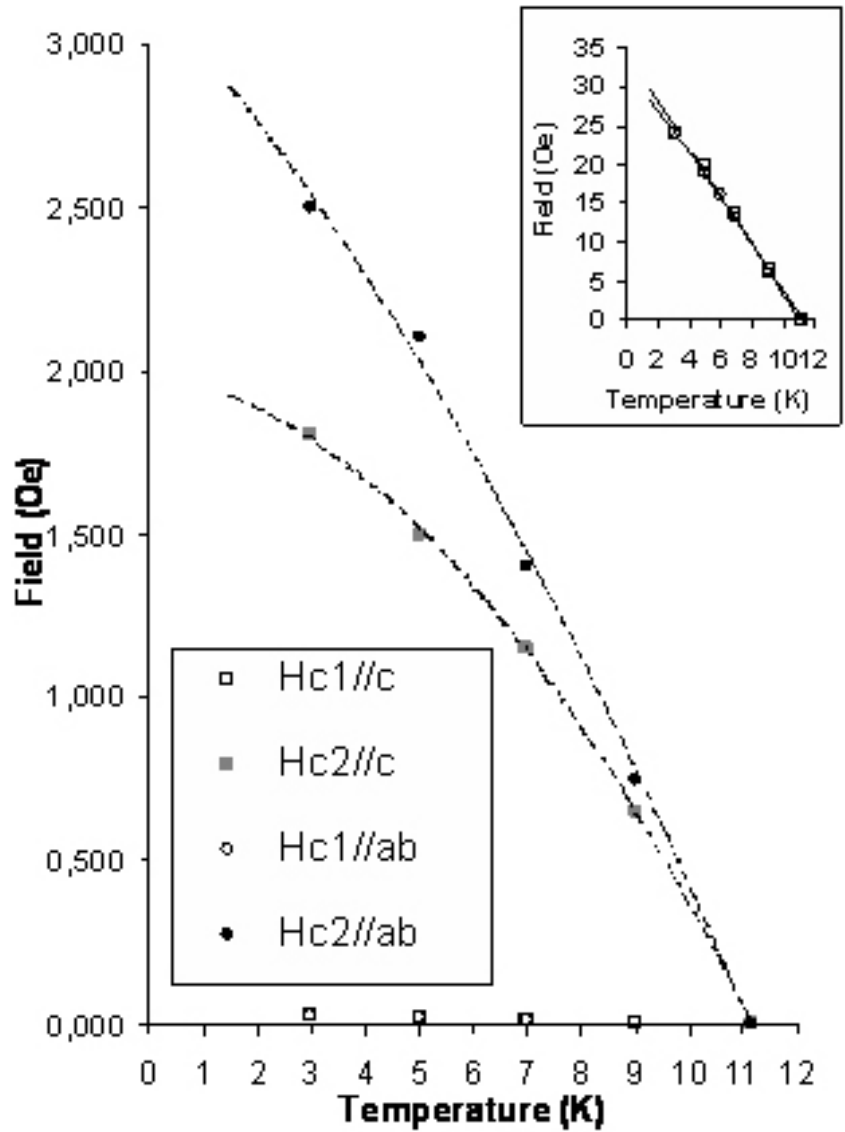

Fig. 3 : Superconducting phase diagram for $Li_3Ca_2C_6$ compiled from M(H) measurements at given temperatures. M(H) data were collected with the external field applied parallel to the **c**-axis ($H_{c1//c}$, $H_{c2//c}$) and in the perpendicular plane ($H_{c1//ab}$, $H_{c2//ab}$).